\documentstyle[sprocl]{article}

\newcommand{\be}{\begin{equation}}
\newcommand{\ee}{\end{equation}}

\begin{document}

\begin{flushright}
EFI-2000-11\\
Fermilab-Conf-00-077-T\\
\end{flushright}
\vskip .2in

\title{ELECTROWEAK SYMMETRY BREAKING BY EXTRA DIMENSIONS}

\author{Hsin-Chia Cheng}

\address{Enrico Fermi Institute, The University of Chicago, Chicago,
 Illinois 60637
}

\author{Bogdan A. Dobrescu, \ Christopher T. Hill}

\address{Theoretical Physics Department, 
Fermilab, Batavia, Illinois 60510
}  


\maketitle

\abstracts{Electroweak symmetry breaking may be naturally induced
by the observed quark and gauge fields in extra dimensions without a
fundamental Higgs field. We show that a composite Higgs doublet can
arise as a bound state of $(t, b)_L$ and a linear combination of the
Kaluza-Klein states of $t_R$, due to QCD in extra dimensions.  The top
quark mass depends on the number of active $t_R$ Kaluza-Klein modes,
and is consistent with the experimental value.  }

Understanding the origin of electroweak symmetry breaking is 
currently one of the most important questions in particle physics.
In the Standard Model (SM), it is assumed that there exists a
fundamental  scalar Higgs field with a negative squared mass, whose
vacuum expectation value (vev) breaks the electroweak symmetry.
However, the mass squared of a fundamental scalar
field receives  a quadratic divergent contribution from radiative
corrections, therefore suffers from the ``hierarchy problem'' if the
cutoff scale is much higher than the weak scale. Recently, a new
solution to the hierarchy problem has been proposed by postulating the
existence of large extra dimensions.\cite{ADD} In that case, the
fundamental scale (which is assumed to be the cutoff scale) in
the higher dimensional theory can be reduced to the TeV range and
hence it removes the large hierarchy between the weak scale and the
Planck scale.  However, just removing the large hierarchy, while
avoiding the fine-tuning problem, does not explain why the electroweak
symmetry is broken, {\it i.e.}, why there is a Higgs field and why its
squared mass is negative.  Here we point out that the extra
dimensions can also provide a natural mechanism for electroweak
symmetry breaking without introducing a fundamental Higgs field.

In fact, a composite Higgs field can arise in the presence of certain
strongly coupled four-quark operators.\cite{Nambu} These four-quark
operators are naturally induced by the Kaluza-Klein (KK) excitations
of the gluons if QCD lives in compact extra dimensions.\cite{ewsb} The
strength of these contact interactions depends on the ratio of the
compactification scale, $M_c$, and the fundumental scale $M_s$ of the
quantum gravitational effects which cuts off the higher  dimensional
gauge interactions. It has been argued that the SM gauge couplings can
unify in the presence of extra dimensions due to the accelerated
power-law running.\cite{DDG} Assuming that they unify at
$M_s$, one typically finds that the gauge couplings
already become strong at that scale for more than two extra
dimensions. Thus, it is possible that the non-perturbative effects form
a composite Higgs out of the quarks.

The simplest 4-dimensional top-condensate model of a composite Higgs
predicts a too large top-quark mass, of approximately $500-600$ GeV, 
if the compositeness
scale is not much higher than the weak scale.\cite{Nambu} For a viable
composite Higgs model, some vector-like quarks are required
to participate in the binding mechanism together with the top
quark.\cite{seesaw} They should have the same SM quantum numbers as
the top quark, so they can naturally be  identified as the KK
excitations of the top quark in a theory with extra
dimensions.\cite{CDH} For instance,  if we assume that only the
right-handed top lives in extra dimensions, the Higgs doublet will be
a bound state of the left-handed top-bottom doublet ($\psi_L$) and a
linear combination of the KK modes of $t_R$,
\be
H\; \sim \psi_L \sum_{i=0}^{n_{\rm KK}-1} \chi_R^i  ,
\ee
where $\chi_R^0=t_R$.  The top quark mass will be suppressed by
$1/\sqrt{n_{\rm KK}}$ compared with the prediction of the  simplest
top condensate model. For a typical $n_{\rm KK}\sim 10-30$, the top
Yukawa coupling is around 1, in agreement with the experimental result.

\begin{figure}[t]
\centering
\begin{picture}(200,100)(0,0)
\put(20,10){\vector(0, 1){70}}
\put(-40,40){\vector(1, 0){300}}
\thicklines
\put(20,40){\line(1, 0){160}}
\put(20,41){\line(1, 0){160}}
\put(20,42){\line(1, 0){160}}
\put(20,60){\line(1, 0){160}}
\put(20,39.5){\line(1, 0){160}}
\put(20,40){\line(0, 1){20}}
\put(180,40){\line(0, 1){20}}
\put(50,37){\large $\times$}  
\put(25,90){$z_{1,...,\delta -1}$}
\put(240,30){$y$}
\put(175,30){\small $L$}
\put(5,60){\small $L_z$}
\put(50,30){\small $y_0$}
\put(20,40){\vector(-1, -1){32}}
\put(-3,4){\small $x^\mu$}
\end{picture}
\caption[]{
\label{Figure1}
\small 
$x^\mu$ are coordinates of the flat four-dimensional spacetime,
$z_1, ..., z_{\delta - 1}$ are the dimensions accessible only to the 
gauge bosons, $\chi$ propagates on  $0\le y\le L$, and 
$\psi_L$ is located at the point marked on the $y$ axis.}
\end{figure}
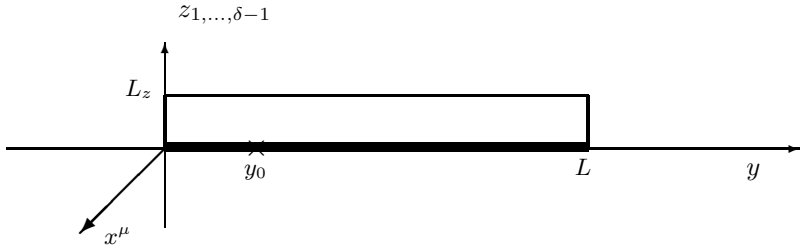

We now present a model for concreteness. This is not a unique choice,
but just an illustrative example of the idea.  We assume that SM gauge
fields propagate in $\delta$ compact  extra dimensions, whose
coordinates are labeled by $y, z_1, ..., z_{\delta-1}$, with sizes
$L$ and $L_z(\ll L)$ respectively.  The $t_R$ is the zero mode
of a 5-dimensional fermion $\chi$, which is fixed at $z=0$, but
propagates on the $[0, L]$ interval in the $y$ direction, while the
$\psi_L=(t, b)_L$ is fixed at $z=0$ and $y=y_0$.  (See
Fig.~\ref{Figure1}.)  The absence of a left-handed zero mode of $\chi$
can result from some boundary condition such as the $S_1/Z_2$ orbifold
projection.  For simplicity, we assume that all other quarks are
4-dimensional fields with left- and right-handed quarks
localized at different positions in extra dimensions so that they do
not form bound states.

Because $L_z \ll L$, we first integrate out the $z$ directions
below the scale $L_z^{-1}$ and obtain a 5-dimensional effective
theory.  The 4-quark interactions are induced by the KK gluons in the
$z$ directions.  After a Fierz transformation, the 4-quark interactions
at the compositeness scale $\Lambda$ are given by
\be
\frac{cg_5^2}{\Lambda^2}
\left\{ \delta(y - y_0)
\left(\overline{\psi}_L \chi\right) \left(\overline{\chi} \psi_L\right)
+ \frac{5}{16}\left[\left(\overline{\chi}\chi\right)^2
- \frac{1}{3} \left(\overline{\chi}\gamma_5\chi\right)^2 \right]
\right\}
+ ... ~,
\label{four}
\ee
where $c \gg 1$ represents the effect of suming over gluon KK modes,
and the ellipsis stand for vectorial and tensorial four-quark
operators, which are not relevant at low energies.

{}From eq.~\ref{four} one can see that the attractive interactions can
give rise to the following bound states: a 4-dimensional weak doublet
complex scalar, $H(x^\mu) \sim \overline{\chi} \psi_L$, and  a
5-dimensional gauge singlet real scalar, $\varphi(x^\mu, y)\sim
\overline{\chi} \chi$. They obtain non-zero kinetic terms in running
down to low energies and become dynamical fields. They also receive
large negative contributions to their squared masses. Electroweak
symmetry is broken when the squared mass of the composite Higgs becomes
negative. Decomposing $\chi$ into 4-dimensional KK states, one can see
indeed that the Higgs field is a bound state of the doublet $\psi_L$
and a linear combination of the KK modes of $t_R$.  Since the observed
(right-handed) top quark is only one of the  $n_{\rm KK}$ states which
participate in the electroweak symmetry breaking, the top Yukawa
coupling is adequately suppressed by  $1/\sqrt{n_{\rm KK}}$ as mentioned 
above.

Whether the singlet $\varphi$ affects the low energy physics at the
weak scale depends on the setup. To illustrate this we consider two
special cases. First, if $\psi_L$ is located at the boundary
($y_0=0$), the Higgs is more strongly bound and hence $\varphi$ will
remain heavy when $m_H^2$ becomes negative.  There is no mixing between
$H$ and $\varphi$ because $\varphi$ vanishes at the boundary. The low
energy theory is simply the Standard Model with a composite Higgs
boson which is expected to be heavy. It can still be consistent with
the electroweak precision measurements because of the mixings between
$t, W, Z$ and their KK states.\cite{precision}

Another interesting case is that $\psi_L$ is localized in the middle
of the $0<y<L$ interval. In this case, the attractive
interactions in the $H$ and $\varphi$ channels are comparable,
therefore both can obtain non-zero vevs. The mixing between $H$ and
$\varphi$ can make either the Higgs boson or the singlet light. If the
mass of the lightest singlet is less than half of the Higgs mass, the
Higgs boson may decay predominantly into two $\varphi$'s, which
subsequently decay into two gluons or two photons through the top
loop, because the Higgs boson interacts strongly with $\varphi$. This
will modify the Higgs search at future colliders.

The model we have summarized here represents a minimal model 
in extra dimensions. The only fields present at the fundamental scale
are the standard fermions and gauge bosons in the higher dimensional 
spacetime. It is remarkable that at energies below the compactification scale
this simple theory reproduces the Standard Model, with the possible addition
of a light singlet scalar. We emphasize though that there is need for 
flavor violating operators at the fundamental scale in order to 
generate masses for the quarks (other than top) and leptons. These 
operators induce Yukawa couplings at low energy. Therefore 
the fermion masses are accomodated as in the SM without 
theoretical predictions other than the top mass.

\end{document}